\documentclass{article}

\begin{document}

\title{Inequalities for generalized parton distributions $H$ and $E$
}
\author{P.V. Pobylitsa}
\date{}
\maketitle

\begin{center}
Institute for Theoretical Physics II, Ruhr University Bochum,\\[0pt]
D-44780 Bochum, Germany \\[0pt]
Petersburg Nuclear Physics Institute, Gatchina, St. Petersburg, 188350,
Russia
\end{center}

\begin{abstract}
New positivity bounds are derived for quark and gluon
generalized parton distributions $H$ and
$E$ in spin 1/2 hadrons.
\end{abstract}

Generalized parton distributions (GPDs)
appear in the QCD description
of a number of hard exclusive processes
\cite{MRGDH-94,Radyushkin-96,Ji-97,CFS-97,Radyushkin-97}.
Since our current knowledge about GPDs is rather poor,
any additional information deserves attention and from this point of
view so called positivity bounds
\cite{Martin-98, Radyushkin-99, PST-99}
for GPDs are of certain interest.

In this paper the following bound for quark generalized parton distributions
$H_{q}$ and $E_{q}$ of spin 1/2 hadrons is derived
\begin{equation}
\left| H_{q}(x,\xi ,t)-\frac{\xi ^{2}}{1-\xi ^{2}}E_{q}(x,\xi ,t)\right|
^{2}+\left| \frac{\sqrt{t_{0}-t}}{2M\sqrt{1-\xi ^{2}}}E_{q}(x,\xi
,t)\right| ^{2}\leq \frac{q(x_{1})q(x_{2})}{1-\xi ^{2}}\,.
\label{new-bound}
\end{equation}
Here the notation of X.~Ji  is used for quark GPDs $H_{q},E_{q}$ and for their arguments
$x,\xi ,t$ (see e.g. \cite{Ji-98}), $M$ is the mass of the hadron.
The usual (forward) unpolarized quark
distributions $q(x)$ are taken at values of $x$

\begin{equation}
x_{1}=\frac{x+\xi }{1+\xi },\quad x_{2}=\frac{x-\xi }{1-\xi }\,.
\label{x-x1-x2-xi}
\end{equation}
Inequality (\ref{new-bound}) is written for the region
$|\xi|<|x|$.
The parameter
\begin{equation}
t_{0}=-\frac{4\xi ^{2}M^{2}}{1-\xi ^{2}}
\end{equation}
corresponds to the maximal value of the squared momentum transfer $t$ ($t\le
t_{0}\le 0$).

Obviously the bound (\ref{new-bound}) is stronger than the following
inequality derived in ref.~\cite{DFJK-01}
\begin{equation}
\left| H_{q}(x,\xi ,t)-\frac{\xi ^{2}}{1-\xi ^{2}}E_{q}(x,\xi ,t)\right|
\leq \sqrt{\frac{q(x_{1})q(x_{2})}{1-\xi ^{2}}}\,,  \label{H-DFJK}
\end{equation}
We also see from inequality (\ref{new-bound}) that
\begin{equation}
\frac{\sqrt{t_{0}-t}}{2M}\left| E_{q}(x,\xi ,t)\right| \leq \sqrt{
q(x_{1})q(x_{2})}\,.  \label{E-DFJK}
\end{equation}
This inequality agrees with the one derived in ref.~\cite{DFJK-01} up to a typo
there \cite{Diehl-communication}.

As noticed in ref.~\cite{DFJK-01} the earlier bound suggested in
refs. \cite{Radyushkin-99,PST-99}
\begin{equation}
\left| H_{q}(x,\xi ,t)\right| \leq \sqrt{\frac{q(x_{1})q(x_{2})}{1-\xi ^{2}}}
\label{H-old}
\end{equation}
is not justified since in its derivation the $E_{q}$ contribution was
overlooked.

The derivation of inequality (\ref{new-bound}) starts from the positivity condition
\cite{Radyushkin-99,PST-99,DFJK-01}
\begin{equation}
\left\| \sum\limits_{k=1}^{2}b_{\lambda }^{k}
\int
\frac{d\tau }{2\pi }e^{i\tau x_{k}(P_{k}n)}\psi _{\alpha }\left( \tau
n\right) |P_{k},\lambda \rangle \right\| ^{2}\geq 0\,.
\label{start-ineq}
\end{equation}
The lhs contains a superposition of two states obtained by acting with quark
fields $\psi $ on a nucleon state $|P_{k},\lambda \rangle $ with momentum
$P_{k}$ and polarization $\lambda $; $n$ is a light-cone vector and the
coefficients $b_{\lambda }^{k}$ are arbitrary. Expanding this squared sum
and summing over the ``good'' spin components $\alpha $ of
the quark fields we obtain

\[
b_{\lambda}^{1\ast }b_{\lambda }^{1}q(x_{1})
+b_{\lambda}^{2\ast }b_{\lambda }^{2}q(x_{2})
\]
\begin{equation}
+\sqrt{1-\xi ^{2}}\left[ b_{\lambda ^{\prime }}^{1\ast }{\cal  H}_{\lambda
^{\prime }\lambda }^{q}(-\Delta )b_{\lambda }^{2}+b_{\lambda ^{\prime
}}^{2\ast }{\cal  H}_{\lambda ^{\prime }\lambda }^{q}(\Delta )b_{\lambda
}^{1}\right] \geq 0\,.  \label{ineq-1}
\end{equation}
The summation over repeated indices is implied, $\Delta =P_{2}-P_{1}$, and
the matrix ${\cal  H}_{\lambda ^{\prime }\lambda }^{q}$ is taken from
ref.~\cite{DFJK-01}
\begin{equation}
{\cal  H}_{\lambda ^{\prime }\lambda }^{q}(\Delta )=\left(
\begin{array}{cc}
H_{q}-\frac{\xi ^{2}}{1-\xi ^{2}}E_{q} & \frac{-\Delta ^{1}+i\Delta ^{2}}{
2M(1-\xi ^{2})}E_{q} \\
\frac{\Delta ^{1}+i\Delta ^{2}}{2M(1-\xi ^{2})}E_{q} & H_{q}-\frac{\xi ^{2}}{
1-\xi ^{2}}E_{q}
\end{array}
\right) _{\lambda ^{\prime }\lambda }\,.
\end{equation}
The inequality (\ref{ineq-1}) should hold for any $b_{\lambda }^{k}$. Taking
special cases $b_{-}^{1}=b_{-}^{2}=0$ or $b_{-}^{1}=b_{+}^{2}=0$ one
reproduces the bounds (\ref{H-DFJK}), (\ref{E-DFJK}).

Now let us consider the case of arbitrary $b_{\lambda }^{k}$. One can
rewrite inequality (\ref{ineq-1}) in the following form

\[
\frac{1}{\sqrt{1-\xi ^{2}}}\left[ b_{\lambda }^{1\ast }b_{\lambda
}^{1}q(x_{1})+b_{\lambda }^{2\ast }b_{\lambda }^{2}q(x_{2})\right] +\left(
b_{\lambda }^{1\ast }b_{\lambda }^{2}+b_{\lambda }^{2\ast }b_{\lambda
}^{1}\right) \left( H_{q}-\frac{\xi ^{2}}{1-\xi ^{2}}E_{q}\right)
\]
\begin{equation}
+i\left( b_{\lambda _{2}}^{1\ast }b_{\lambda _{1}}^{2}-b_{\lambda
_{2}}^{2\ast }b_{\lambda _{1}}^{1}\right) \left( \Delta ^{1}\sigma
^{2}-\Delta ^{2}\sigma ^{1}\right) _{\lambda _{2}\lambda _{1}}\frac{E_{q}}{
2M(1-\xi ^{2})}\geq 0
\end{equation}
where $\sigma ^{k}$ are Pauli matrices. Rotating the spinor indices $\lambda $
of $b_{\lambda }^{k}$ by an $SU(2)$ transformation one can diagonalize
$\Delta ^{1}\sigma ^{2}-\Delta ^{2}\sigma ^{1}\rightarrow |\Delta _{\perp
}|\sigma ^{3}$ so that the inequality takes the form
\[
\frac{1}{\sqrt{1-\xi ^{2}}}\left[ b_{\lambda }^{1\ast }b_{\lambda
}^{1}q(x_{1})+b_{\lambda }^{2\ast }b_{\lambda }^{2}q(x_{2})\right]
\]
\[
+\left( b_{\lambda }^{1\ast }b_{\lambda }^{2}+b_{\lambda }^{2\ast
}b_{\lambda }^{1}\right) \left( H_{q}-\frac{\xi ^{2}}{1-\xi ^{2}}E_{q}\right)
\]
\begin{equation}
+i\left( b_{\lambda _{2}}^{1\ast }b_{\lambda _{1}}^{2}-b_{\lambda
_{2}}^{2\ast }b_{\lambda _{1}}^{1}\right) |\Delta _{\perp }|\sigma _{\lambda
_{2}\lambda _{1}}^{3}\frac{E_{q}}{2M(1-\xi ^{2})}\geq 0\,.
\end{equation}
We see that we have two independent inequalities for $b_{+}^{k}$ and $
b_{-}^{k}$
\begin{equation}
\sum\limits_{k,k'}
b_{\pm }^{k^{\prime }}\left(
\begin{array}{cc}
\frac{q(x_{1})}{\sqrt{1-\xi ^{2}}} & H_{q}-\frac{\xi ^{2}}{1-\xi ^{2}}
E_{q}\pm i\frac{|\Delta _{\perp }|E_{q}}{2M(1-\xi ^{2})} \\
H_{q}-\frac{\xi ^{2}}{1-\xi ^{2}}E_{q}\mp i\frac{|\Delta _{\perp }|E_{q}}{
2M(1-\xi ^{2})} & \frac{q(x_{2})}{\sqrt{1-\xi ^{2}}}
\end{array}
\right) _{k^{\prime }k}b_{\pm }^{k}\geq 0\,.
\end{equation}

The positivity of this quadratic form immediately leads us to the constraint
\begin{equation}
\left( H_{q}-\frac{\xi ^{2}}{1-\xi ^{2}}E_{q}\right) ^{2}+\left( |\Delta
_{\perp }|\frac{E_{q}}{2M(1-\xi ^{2})}\right) ^{2}\leq \frac{q(x_{1})q(x_{2})
}{1-\xi ^{2}}\,.
\end{equation}
Taking into account that
\begin{equation}
t=\frac{4\xi ^{2}M^{2}+|\Delta _{\perp }|^{2}}{1-\xi ^{2}}
\end{equation}
one arrives at the result (\ref{new-bound}).

Note that excluding $E_{q}$ from the bound (\ref{new-bound}) we obtain a
constraint on $H_{q}$
\begin{equation}
|H_{q}(x,\xi ,t)|\leq \sqrt{\left( 1-\frac{t_{0}}{t}\right) ^{-1}\frac{
q(x_{1})q(x_{2})}{1-\xi ^{2}}}\,.  \label{H-only}
\end{equation}
Since $t\leq t_{0}\leq 0$ this bound is weaker than the obsolete bound
(\ref{H-old}).

In a similar way one can derive the following bound for the gluon GPDs
$H_{g},$ $E_{g}$ (normalized according to
the convention of ref.~\cite{Diehl-0101335})
\[
\left| H_{g}(x,\xi ,t)-\frac{\xi ^{2}}{1-\xi ^{2}}E_{g}(x,\xi ,t)\right|
^{2}
\]
\begin{equation}
+\left| \frac{\sqrt{t_{0}-t}}{2M\sqrt{1-\xi ^{2}}}E_{g}(x,\xi ,t)\right|
^{2}\leq \frac{x^{2}-\xi ^{2}}{1-\xi ^{2}}g(x_{1})g(x_{2})
\label{H-E-g}
\end{equation}
where $g(x)$ is the forward gluon distribution. In particular, this allows us
to
to reproduce the following result from ref.~\cite{DFJK-01}
\begin{equation}
\left| H_{g}(x,\xi ,t)-\frac{\xi ^{2}}{1-\xi ^{2}}E_{g}(x,\xi ,t)\right|
\leq \sqrt{\frac{x^{2}-\xi ^{2}}{1-\xi ^{2}}g(x_{1})g(x_{2})}\,.
\end{equation}

By analogy with (\ref{H-only}) we find from (\ref{H-E-g})
\begin{equation}
|H_{g}(x,\xi ,t)|\leq \sqrt{\left( 1-\frac{t_{0}}{t}\right) ^{-1}
\frac{x^{2}-\xi ^{2}}{1-\xi ^{2}}
g(x_{1})g(x_{2})}\,.
\end{equation}

Although the inequalities derived here are stronger than the bounds
on GPDs found earlier, one should not consider the new results as the
best final
estimates. Indeed, the derivation included the summation over quark polarization
indices in the lhs of (\ref{start-ineq}) and some information could be lost
at this step.

I appreciate discussions with A.~Belitsky, J.C.~Collins, M.~Diehl, N.~Kivel,
D.~M{\"u}ller, V.Yu.~Petrov, M.V.~Polyakov, A.V.~Radyushkin,
M.~Strikman and O.~Teryaev.

\end{document}